\documentclass[11pt,letterpaper,english]{revtex4}
\usepackage{fourier}
\usepackage{avant}
\usepackage[T1]{fontenc}
\usepackage[latin9]{inputenc}
\usepackage{amsthm}
\usepackage{amsmath}
\usepackage{graphicx}
\usepackage{esint}

\makeatletter

\pdfpageheight\paperheight
\pdfpagewidth\paperwidth

\providecommand{\tabularnewline}{\\}

\@ifundefined{textcolor}{}
{%
 \definecolor{BLACK}{gray}{0}
 \definecolor{WHITE}{gray}{1}
 \definecolor{RED}{rgb}{1,0,0}
 \definecolor{GREEN}{rgb}{0,1,0}
 \definecolor{BLUE}{rgb}{0,0,1}
 \definecolor{CYAN}{cmyk}{1,0,0,0}
 \definecolor{MAGENTA}{cmyk}{0,1,0,0}
 \definecolor{YELLOW}{cmyk}{0,0,1,0}
 }

\makeatother

\usepackage{babel}

\begin{document}

\title{Complexity in human transportation networks: A comparative analysis
of worldwide air transportation and global cargo ship movements}

\author{Grastivia O'Danleyman}

\email{grastivia.odanleyman@gmail.com}

\affiliation{Department of Engineering Sciences and Applied Mathematics, Northwestern
University, Evanston, Illinois, USA.}

\author{Jake Jungbin Lee}

\affiliation{Department of Engineering Sciences and Applied Mathematics, Northwestern
University, Evanston, Illinois, USA.}

\author{Hanno Seebens}

\affiliation{ICBM, University of Oldenburg, 26111 Oldenburg, Germany}

\author{Bernd Blasius}

\affiliation{ICBM, University of Oldenburg, 26111 Oldenburg, Germany}

\author{Dirk Brockmann}

\affiliation{Department of Engineering Sciences and Applied Mathematics \& Northwestern
Institute on Complex Systems, Northwestern University, Evanston, Illinois,
USA }
\begin{abstract}
We present a comparative network theoretic analysis of the two largest
global transportation networks: The worldwide air-transportation network
(WAN) and the global cargoship network (GCSN). We show that both networks
exhibit striking statistical similarities despite significant differences
in topology and connectivity. Both networks exhibit a discontinuity
in node and link betweenness distributions which implies that these
networks naturally segragate in two different classes of nodes and
links. We introduce a technique based on effective distances, shortest
paths and shortest-path trees for strongly weighted symmetric networks
and show that in a shortest-path-tree representation the most significant
features of both networks can be readily seen. We show that effective
shortest-path distance, unlike conventional geographic distance measures,
strongly correlates with node centrality measures. Using the new technique
we show that network resilience can be investigated more precisely
than with contemporary techniques that are based on percolation theory.
We extract a functional relationship between node characteristics
and resilience to network disruption. Finally we discuss the results,
their implications and conclude that dynamic processes that evolve
on both networks are expected to share universal dynamic characteristics.
\end{abstract}
\maketitle

\section{Introduction}

Large-scale, human transportation networks are essential for global
travel, international trade, the facilitation of international partnerships
and relations, and the advancement of science and commerce. The worldwide
air transportation network supports the traffic of over three billion
passengers travelling between more than $4,000$ airports on more
than $50$ million flights in a year~\cite{WAN:Online}. The worldwide
cargo ship network accounts for up to 90\% of the international exchange
of goods; approximately 60,000 cargo ships are connecting more than
5000 ports world wide with about a million ship movements every year~\cite{UNCTAD2008,IHSFairplay}.
These two networks constitute the operational backbone of our globalized
economy and society.

Although they are immensely important for facilitating exchange between
geographically distant regions, the ever-increasing amount of traffic
over such complex, densely-connected transportation networks introduces
serious problems. Rising energy costs, pollution, and global warming
are obvious concerns, but globalized traffic also plays a key role
in the worldwide dissemination of infectious diseases and invasive
species~\cite{Colizza:2007p3898,Colizza:2007p1066,Colizza:2006p516,Colizza:2006p1227,Balcan:2009p5179,Kaluza:2010p5210,Tompkins:2003p606,Ruiz:2000p1231,Drury:2007p6024,Ferguson:2006p509,Halloran:2008p1218,Hollingsworth:2006p1132,Hollingsworth:2007p1078,Meyerson2007,Hulme2009,Levine2003}.
The first decade of the 21st century has witnessed the emergence and
worldwide spread of two major global epidemics: the severe acute respiratory
syndrome (SARS) in 2003~\cite{Hufnagel:2004p541,Brockmann:2008p1199,Colizza:2007p1229,Cooper:2006p1090},
and the recent H1N1 pandemic of 2009~\cite{Balcan:2009p5173,Fraser:2009p5194}.
Both diseases rapidly spread across the globe in a matter of weeks
to months, a process linked directly to long-distance traffic routes
over which infected individuals dispersed infectious agents. In combination
with increasing worldwide population size, which is expected to pass
the 7 billion threshold within the next decade, and the concentration
of the majority of the world's population in mega-cities and urban
areas~\cite{UnitedNations:Online}, the impact of global pandemic
events is expected to become one of the most challenging problems
of the 21st century. The spread of invasive species into new habitats
and ecosystems presents a similar and equally-significant problem~\cite{Mack:2000p6063,Kolar2002,Simberloff2005,Meyerson2007}.
The largest vector of marine bioinvasion is assumed to be global shipping~\cite{Ruiz1997}.
Human-mediated bioinvasion has become one of the key factors in the
global biodiversity crisis~\cite{Sala2000,Molnar2008} and may affect
the stability of ecosystems, survival of species, and human health~\cite{Mack:2000p6063,Ruiz:2000p1231}.
The introduction of invasive species to foreign ecosystems has generated
annual costs of over \$120 billion in the United States alone~\cite{Pimentel:2005p6070}. 

In addition to introducing environmental problems, the complex transportation
web itself is subject to external disruptions. For instance, the unexpected
eruption of the Icelandic volcano Eyjafjallajökull in 2010 and subsequent
closure of major European airports led to a major disruption in global
traffic and significant economic stress over a period of several days.
Other influences, such as meteorological events like hurricanes, the
recent rise in acts of piracy in Somalia, or the financial crisis
of 2007 also make flexibility in worldwide cargo traffic necessary
and underline the vulnerabilty of international trade and transportation
systems. It is therefore of fundamental importance to understand the
resilience of these networks in response to regional and large-scale
failure of parts of the network, and to identify {}``sensitive''
regions of the network. This point becomes even more important in
light of malicious terrorist activities.

A deep understanding of the structure of human transportation networks
will lead to new insights into the geographical spread of diseases
and invasive species, allow the development of new computational models
for their time courses, and eventually allow us to predict their impact
on our environment and society. Computational techniques for investigating
the resilience of these networks in the face of partial failure will
play a fundamental part in achieving this understanding; complex network
theory~\cite{Newman:2003p216} already provides a powerful theoretical
tool in this respect. But, although both the worldwide air transportation
network and the global cargo ship network have already been subjected
to a number of network-theoretic analyses~\cite{Barrat:2005p632,Brockmann:2008p1199,Colizza:2006p516,Vespignani:2009p2700,Kaluza:2010p5210},
it is still unclear whether the observed properties of these networks
are unique to a specific context, or are universal and generic. A
lack of comparative studies in this direction, and indeed a lack of
data, has lead to a scarcity of universal theories of the structure
of transportation networks.

Here we address this issue using a comparative approach. We analyse
and compare the structure of the worldwide air-transportation and
cargo-ship networks (WAN and GCSN in the following) and show that
a surprising number of properties are shared by both networks despite
their different use, economic context, scale, and connectivity structures.
The analysis suggests that the same fundamental principles guide the
growth of both networks. Most importantly, it suggests that dynamic
processes that evolve on these networks will exhibit similar dynamic
features, an important insight since it implies that processes as
different as emergent human infectious diseases and human-mediated
bioinvasion can be investigated along the same line of research.

The dynamics of processes on networks are guided not only by the topology
of the network, but also by the interaction strengths between pairs
of nodes. One of the characteristic features of transportation networks
is a strongly heterogeneous distribution of interaction strength.
Among other things, this implies that the shortest topological path
between two nodes may not be the path of strongest interaction, and
we account for such effects by using the idea of effective shortest
paths. These are analogous to the well-known topological shortest
paths, except that the length of an edge is taken to be the reciprocal
of the weight of that edge, and the effective shortest path is then
the path that minimizes the total effective distance. This approach
also reveals surprising similarities between the two networks.

The paper is structured as follows: In Sections~\ref{sec:thenetworks}
and~\ref{sec:Statistics} we introduce the WAN and GCSN and discuss
their statistical properties and similarities. In Section~\ref{sec:reilience}
we investigate and compare resilience of these networks in response
to targeted attacks and random failure. In Section~\ref{sec:SPD_SPTD}
we apply a recently developed technique based on shortest-path trees
to compute the structural backbones of these networks. We discuss
the implications of our results in Section~\ref{sec:discussion}.

\section{The worldwide air transportation and cargo ship networks}

\label{sec:thenetworks}%
\begin{figure}
\includegraphics[width=1\columnwidth]{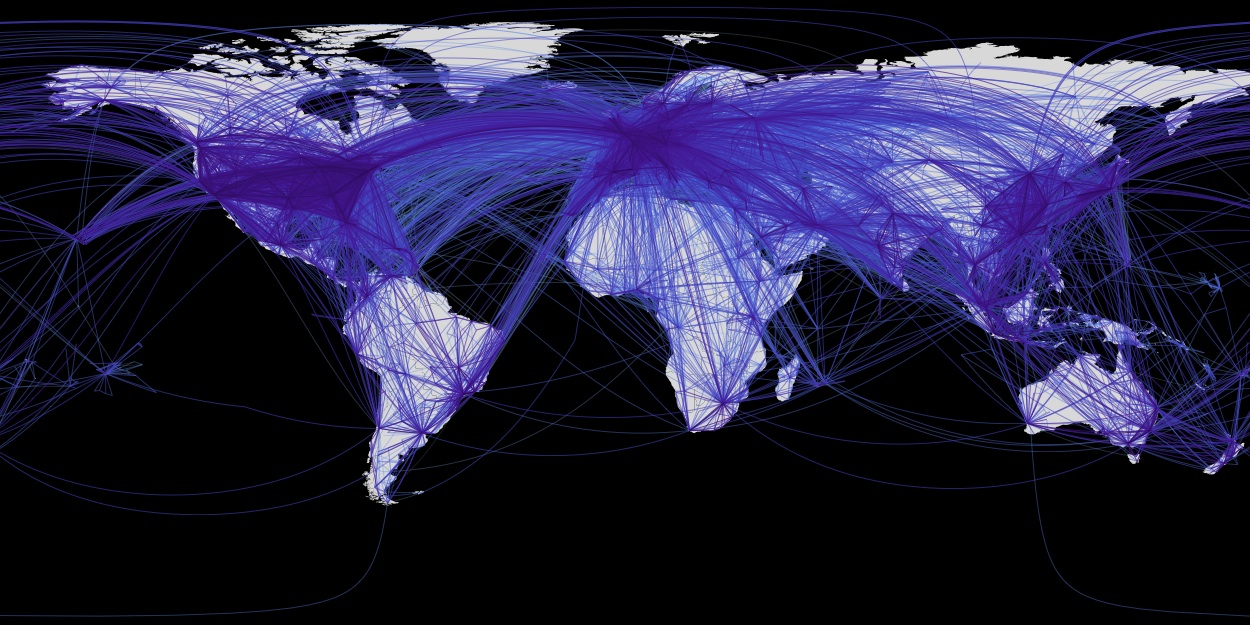}

\includegraphics[width=1\columnwidth]{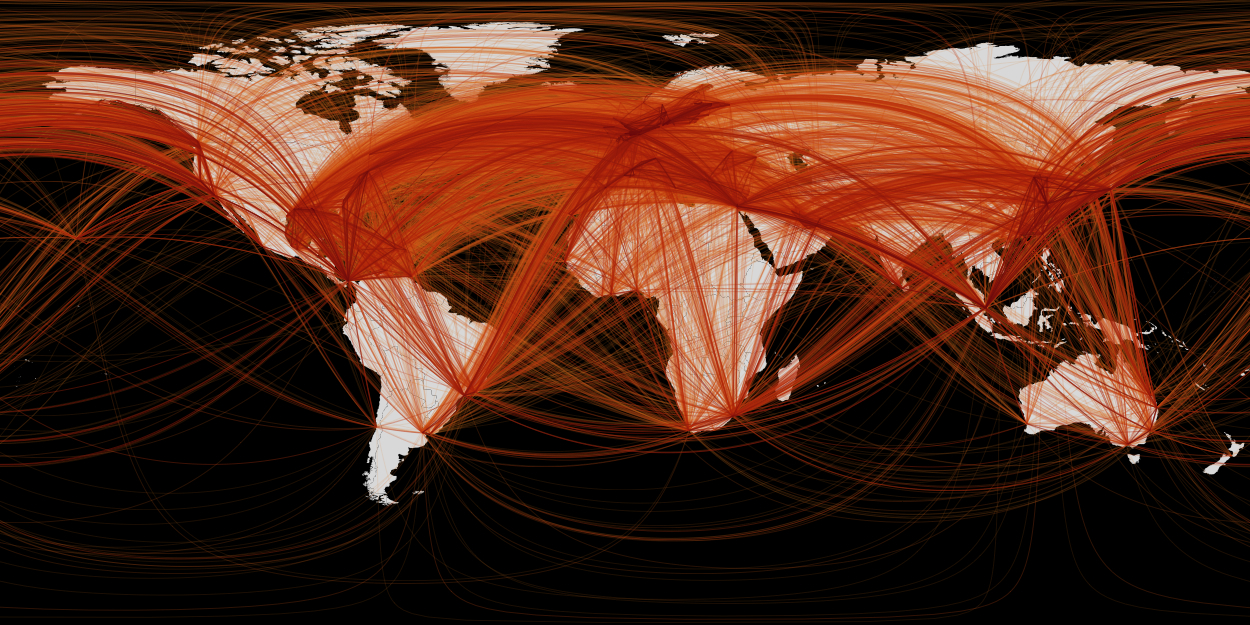}

\caption{\label{fig:thenetworks}Global Transportation and Mobility: The top
network illustrates the worldwide air-transportation network (WAN)
consisting of $4,069$ airports connected by $25,453$ links that
represent the number of passenger planes travelling between nodes
per unit time. The saturation of the lines indicates the total flux
along a route. The bottom panel depicts the global cargo-ship network
(GCSN) that connects $951$ international ports along $25,819$ routes.}
\end{figure}
The WAN and GCSN are infrastructure systems on which we travel and
transport commodities on a worldwide scale. Complex network theory~\cite{Newman:2003p216,Barrat:2004p4820,DallAsta:2006p3878}
provides the most plausible quantitative description of these systems:
pairs of nodes $i$ and $j$ are connected by links with non-negative
weights $w_{ij}>0$ if transport occurs directly between these nodes,
and $w_{ij}=0$ if they are not directly connected. It is generally
possible in transportation networks to begin at any node and locate
a path to any other node, but the $w_{ij}$ measure only direct connections
and quantify the magnitude of traffic between pairs of nodes. In the
WAN nodes represent airports and the weight matrix could be defined
as the total number of passengers per unit time, the number of passenger
planes or the number of scheduled flights. In the GCSN nodes represent
ports, and $w_{ij}$ could quantify the number of cargo ships or the
net tonnage of cargo per unit time. For a comparative analysis we
choose $w_{ij}$ to represent the number of carrier vehicles (passenger
planes or cargo ships) that travel from node $j$ to $i$ per unit
time. For the WAN, $w_{ij}$ is the average number of scheduled commercial
flights per year between airports in the three-year period 2004--2006,
as reported by OAG Worldwide Ltd.~\cite{WAN:Online}. The GCSN was
established from data of world-wide ship movements provided by IHS
Fairplay~\cite{IHSFairplay}. This information was used to reconstruct
the journey of 15,415 ships travelling around the globe during 2007~\cite{Kaluza:2010p5210}.
The GCSN is restricted to vessels bigger than 10,000~gross~tonnages
which accounts for 86\% of the world fleet. For both, WAN and GCSN,
the resulting weight matrix $W$ is virtually symmetric, i.e.~$w_{ij}\approx w_{ji}$
for all pairs $ij$. In order to guarantee strict symmetry we symmetrized
the matrix according to $w_{ij}\rightarrow(w_{ij}+w_{ji})/2$. This
differs from the definition that was used in ~\cite{Kaluza:2010p5210},
where the number of directed links was reported.

Both networks are depicted in Fig.~\ref{fig:thenetworks}. %
\begin{table}
\begin{tabular}{|c|c|c|c|c|c|c|c|c||c|c|c||c|c|c|}
\hline 
 & $N$ & $L$ & $\sigma$ & $\phi$ & $d_{\text{T}}$ & $c$ & $\left\langle r\right\rangle $ & $C$ & $\left\langle w\right\rangle $  & $\left\langle F\right\rangle $ & $\left\langle k\right\rangle $ & $\text{CV}(w)$ & $\text{CV}(F)$ & $\text{CV}(k)$\tabularnewline
\hline
\hline 
WAN & $4,069$ & $25,453$ & $3.07\times10^{-3}$ & $29.11$ & $4.16$ & $0.55$ & $1,109$ & $5.68\times10^{7}$ & $1116.49$ & $1.39\times10^{4}$ & $12.51$ & $2.03$ & $3.60$ & $2.15$\tabularnewline
\hline 
GCSN & $951$ & $25,819$ & $5.72\times10^{-2}$ & $13.57$ & $2.34$ & $0.57$ & $1,857$ & $9.87\times10^{5}$ & $19.11$ & $1.04\times10^{3}$ & $54.30$ & $7.27$ & $2.34$ & $1.22$\tabularnewline
\hline
\end{tabular}

\caption{\label{tab:networkproperties}Network characteristics of WAN and GCSN.
Number of nodes $N$, number of links $L$, network connectivity $\sigma\approx2L/N^{2}$,
network diameters $\phi$ and $d_{\text{T}},$ network clustering
coefficient $c$ and network length scale $\left\langle r\right\rangle $
(in units km), total traffic $C$ (vehicles/yr.) reflect global properties
of the network. The table also lists the mean link weights $\left\langle w\right\rangle $,
traffic per node (flux) $\left\langle F\right\rangle $ (both in units
of vehicles/yr.) and node degree $\left\langle k\right\rangle $ and
corresponding coefficients of variation.}
\end{table}
Despite their global coverage and structural similarity, these networks
exhibit distinct features. The WAN comprises approximately five times
as many nodes but almost the same number of links, yielding a less-densely-connected
network ($\sigma_{\text{WAN}}=3.07\times10^{-3}$ as compared to $\sigma_{\text{GCSN}}=5.72\times10^{-2}$,
see Table~\ref{tab:networkproperties}). Note that the GCSN here
is restricted to large vessels~\cite{Kaluza:2010p5210} and consequently
the total number of seaports in the world may be an order of magnitude
higher, and thus in a similar range as the WAN.

Node flux and degree are key characteristics, defined according to\begin{equation}
F_{i}=\sum_{j}w_{ij}\quad\text{and}\quad k_{i}=\sum_{j}a_{ij},\label{eq:flux and degree}\end{equation}
where $a_{ij}$ are elements of the adjacency matrix, that is $a_{ij}=1$
if nodes $i$ and $j$ are connected and $a_{ij}=0$ otherwise. On
average a node in the WAN dispatches $\left\langle F\right\rangle _{\text{WAN}}=1.39\times10^{4}$
vehicles per year. The average number of cargo ships leaving a port
in the GCSN is $\left\langle F\right\rangle _{\text{GCSN}}=1.03\times10^{4}.$
Higher connectivity of the GCSN is also reflected in the mean degree,
$\left\langle k\right\rangle _{\text{WAN}}=12.51$ and $\left\langle k\right\rangle _{\text{GCSN}}=54.30$.

The typical traffic per link is given by the mean link weight $\left\langle w\right\rangle $,
and although in the WAN it exceeds the GCSN by two orders of magnitude,
the variability reflected in the coefficient of variation is significantly
higher in the GCSN. The clustering coefficient $c$ indicates the
abundance of triangular motifs in the network, and in spite of the
GCSN's higher connectivity the clustering coefficient is nearly identical
in both networks. This indicates that both networks can be considered
sparse and saturation effects are not significant.

A typical length scale of the network can be defined by \begin{equation}
\left\langle r\right\rangle =\sum_{j,i}p_{ji}r_{ji}\label{eq:accessarea}\end{equation}
where $p_{ji}=w_{ji}/F_{j}$ and $r_{ji}$ the geographical distance
between nodes $j$ and $i$. The quantity $p_{ji}$ is the relative
fraction of traffic from $i$ to $j$ with respect to the entire traffic
through node $i$. Thus $\left\langle r\right\rangle $ represents
the mean distance traveled by a carrier in the network. According
to this definition the typical length scale of the GCSN is approximately
twice the size of the typical length scale of the WAN, see Table~\ref{tab:networkproperties}.
Related to the geographic distance are topological distance measures
defined by the connectivity of the networks. The diameter of a network
can be defined as the average shortest-path length that connects a
pair of nodes, $d_{\text{T}}$. For WAN and GCSN $d_{\text{T}}=4.16$
and $2.34$, respectively (in a fully connected network $d_{\text{T}}=1$).

\section{Universal statistics in large-scale transportation networks}

\label{sec:Statistics}
\begin{figure}
\includegraphics[width=0.32\columnwidth]{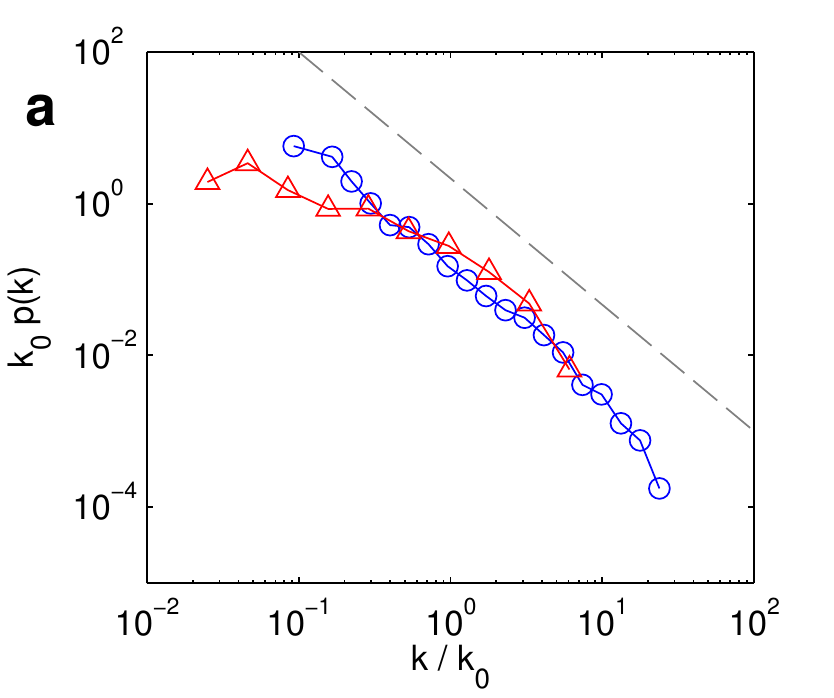}\hfill{}\includegraphics[width=0.32\columnwidth]{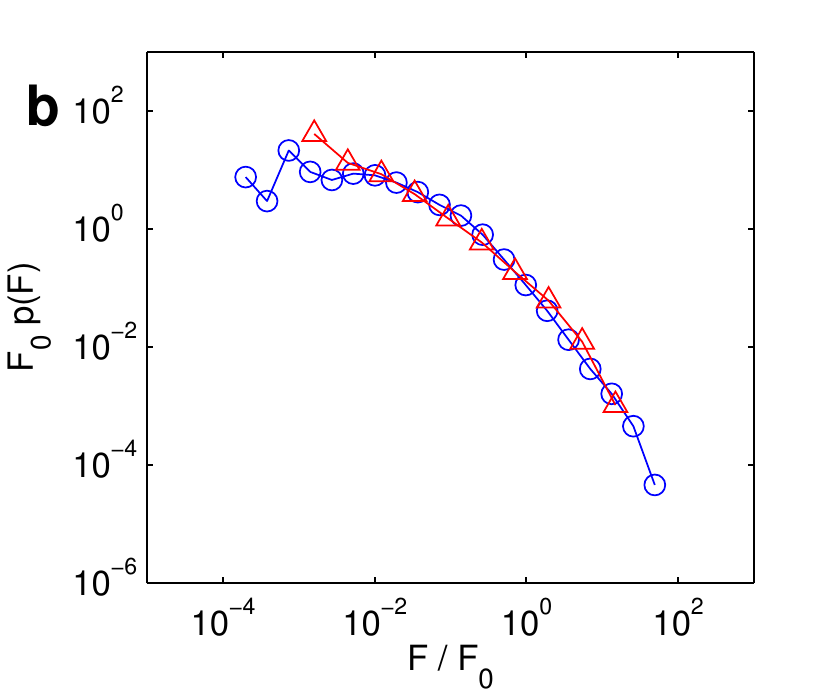}\hfill{}\includegraphics[width=0.32\columnwidth]{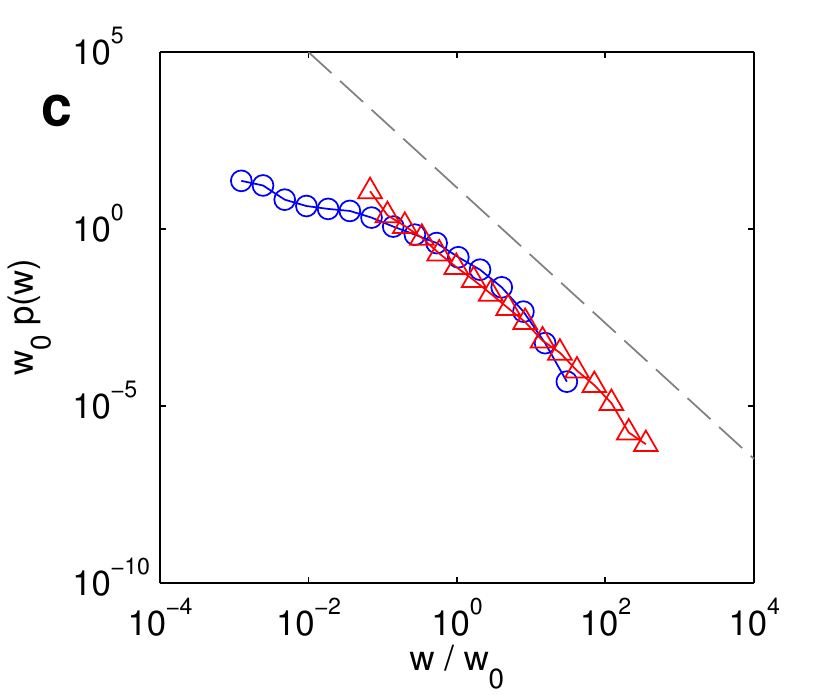}

\caption{\label{fig:network_stats}Statistical properties of WAN (blue) and
GCSN (red): Panels \textbf{a},\textbf{b},\textbf{c} depict the probability
density functions of node degree $k$, node flux $F$ and link weight
$w$. Up to scaling factors these distributions exhibit very similar
functional shapes. Link statistics are shown in \textbf{d} (link weights
$w$) and \textbf{e} (weighted link betweenness $b$). Approximate
scaling behavior is indicated by dashed lines in each panel. Scaling
exponents in \textbf{a} and \textbf{c} are $\alpha=1.5$ and $\beta=2$,
respectively. The abscissal scaling factors $k_{0},F_{0},w_{0}$ are
given by the mean in each distribution.}
\end{figure}

A key feature that many large-scale technological
networks share is their strong structural heterogeneity in terms of
link and node statistics and centrality measures. These networks typically
contain a small fraction of hubs characterized by strong connectivity
and high centrality scores complemented by a large number of smaller
nodes that connect to the hubs. Distributions of centrality measures
are often scale-free~\cite{Barabasi:1999p3892,Vespignani:2009p2700}.
Both the WAN and the GCSN exhibit these structural properties and,
more importantly, their centrality statistics are almost identical,
as is illustrated in Fig.~\ref{fig:network_stats}, which shows the
relative frequencies of link weights, node degree, and node flux.

Figure~\ref{fig:network_stats} suggests that $w$, $k$, and $F$
follow almost identical distributions (up to a scaling factor) and
range across many orders of magnitude. Their surprisingly similar
shape supports the claim that these networks have evolved according
to similar fundamental processes. It has been pointed out~\cite{Brockmann:2006p3864,Brockmann:2008p1135,DallAsta:2006p3878,Barrat:2005p632}
that degree, flux, and weight approximately follow power laws. This
is confirmed for $p(w)$ and $p(k)$, and we find\begin{equation}
p(k)\sim k^{-\alpha}\quad\text{and}\quad p(w)\sim w^{-\beta}\label{eq:distributions}\end{equation}
with exponents $\alpha\approx1.5$ and $\beta\approx2$.

\subsection{Weighted betweenness centrality of links and nodes}

\begin{figure}
\hfill{}\includegraphics[width=0.32\columnwidth]{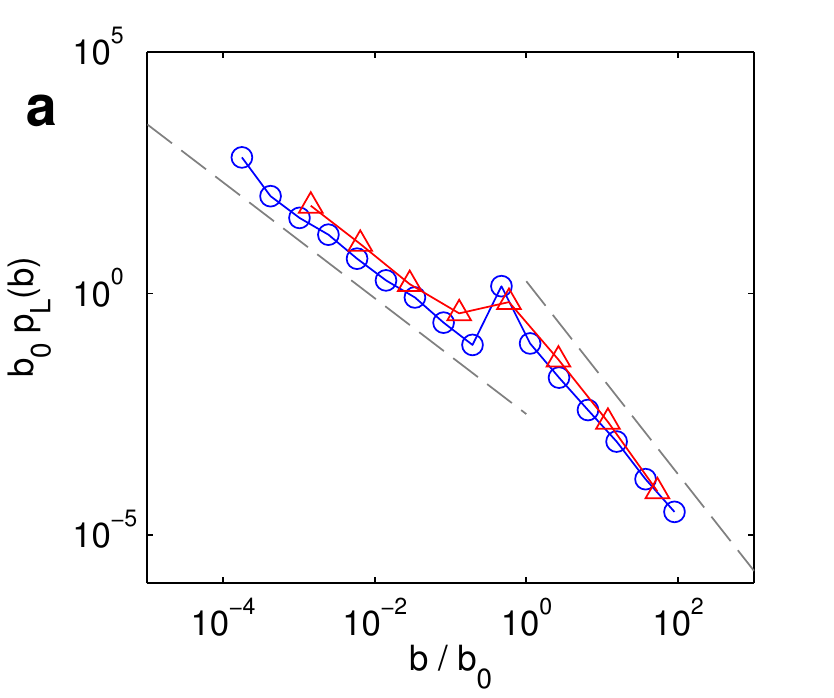}\hfill{}\includegraphics[width=0.32\columnwidth]{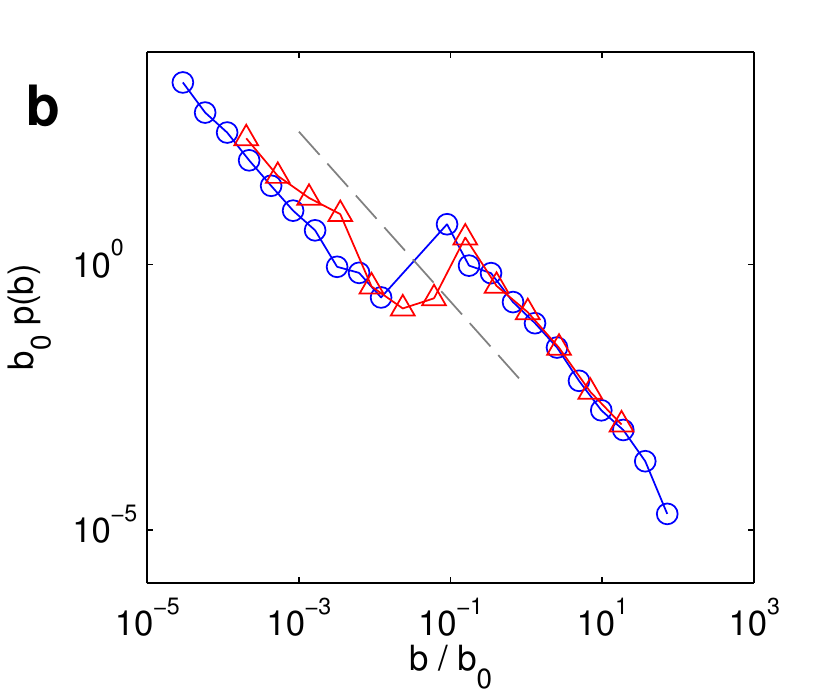}\hfill{}

\caption{\label{fig:network_betweenness}Structure of betweenness centrality
of WAN (blue) and GCSN (red): Panels \textbf{a} and \textbf{b} depict
the distributions $p(b)$ of betweenness of links and nodes, respectively.
Node betweenness exhibits two distinct regimes separated by a distinct
discontinuity for intermediate betweenness values. Each regime is
characterized by a scaling exponent $\gamma_{1}=1.6$. Link betweenness
exhibits two scaling regimes separated by a marked discontinuity as
well, with two different exponents $\gamma_{2}=1.2$ ($b\ll b_{0}$)
and $\gamma_{2}=2.0$ ($b\gg b_{0}$). The abscissal scaling factors
$b_{0}$ are given by the mean in each distribution.}
\end{figure}
 Another commonly investigated measure for link and node centrality
is betweenness centrality. The betweenness $b$ of a link (or a node)
is the fraction of shortest paths in the entire network of which the
link (or node) is part of. Betweenness requires the definition of
length of a path which in turn requires the definition of length of
a link. In weighted networks a plausible choice for the effective
length of a link connecting nodes $i$ and $j$ is given by the proximity
$\lambda_{ij}$ defined by

\begin{equation}
\lambda_{ij}=\frac{\left\langle w\right\rangle }{w_{ij}}.\label{eq:proximity}\end{equation}
This definition accounts for the notion that strongly connected nodes
are effectively more proximate than nodes that are weakly coupled.
The numerator $\left\langle w\right\rangle $ sets the typical distance
scale $\lambda_{0}=1/\left\langle w\right\rangle $ and $\lambda_{ij}$
is defined relative to it. Based on this effective proximity one can
define the length of a path $P(i_{0},...,i_{k})$ that starts at node
$i_{0}$ and terminates at node $i_{k}$ connecting a sequence of
intermediate nodes $i_{n}$, $n=1,...,k-1$ along direct connections
of weights $w_{i_{n}i_{n+1}}$ by summing of the proximities of each
leg in the path. This integrated distance $l(i_{0},...i_{k})$ is
given by the sum \begin{equation}
l(i_{0},...i_{k})=\sum_{n=0}^{k-1}\lambda_{i_{n}i_{n+1}}=\sum_{n=0}^{k-1}\frac{\left\langle w\right\rangle }{w_{i_{n}i_{n+1}}}.\label{eq:length of path}\end{equation}
For a given pair of nodes, $i_{0}$ and $i_{k}$ many paths exists
that connect these nodes along intermediate nodes $i_{n}=1,...,k-1$.
Using the definition of length of a path above, the shortest path
between two nodes is defined as one with minimal $l$.\[
d(i_{0},i_{k})=\min_{i_{n}=1,..k-1}l(i_{0},...,i_{k}).\]
We define the effective distance $d_{ij}$ between nodes $i$ and
$j$ as this effective length of the shortest path connecting them,
i.e. $d_{ij}=d(i,j)$ and denote the unique path associated with it
by $P_{s}(i,j)$. Based on this definition we define the diameter
$\phi$ of the network as the mean shortest-path length over the ensemble
of all pairs of nodes. According to this definition the WAN's diameter
is slightly more than twice the diameter of the GCSN, see Table~\ref{tab:networkproperties}.
The reasons for this will be discussed in more detail in Section~\ref{sec:SPD_SPTD}. 

We computed betweenness centrality $b$ for both, links and nodes
based on the set of all shortest paths $P(i,j)$. Figure~\ref{fig:network_betweenness}
depicts the distributions $p(b)$ for both networks. Unlike the centrality
measures of degree and flux for nodes and weights for links, the distribution
of betweenness exhibits a well pronounces discontinuity in both networks.
This indicates that in the WAN and GCSN links and nodes segragate
into two distinct functional groups. In fact the point $b_{c}$ at
which the discontinuity occurs can be employed to separate links and
nodes that belong to the operational backbone of the network~\cite{salience}.
A key observation is that the distribution of betweenness in both
networks is very similar: both exhibit the discontinuity and both
exhibiting scaling behavior in the two betweenness regimes.

\subsection{Correlations in centrality measures}

\begin{figure}
\begin{centering}
\includegraphics[width=0.32\columnwidth]{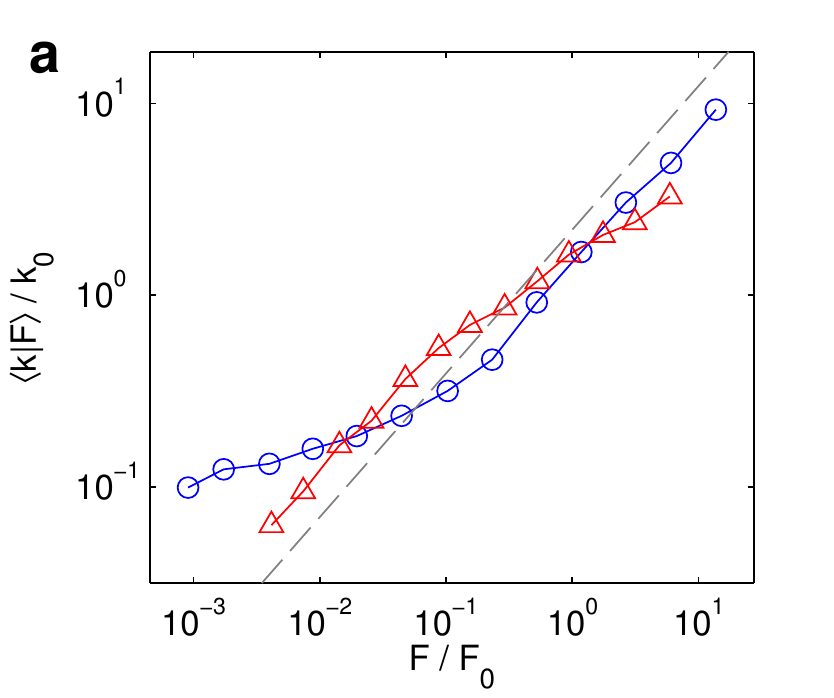}\hfill{}\includegraphics[width=0.32\columnwidth]{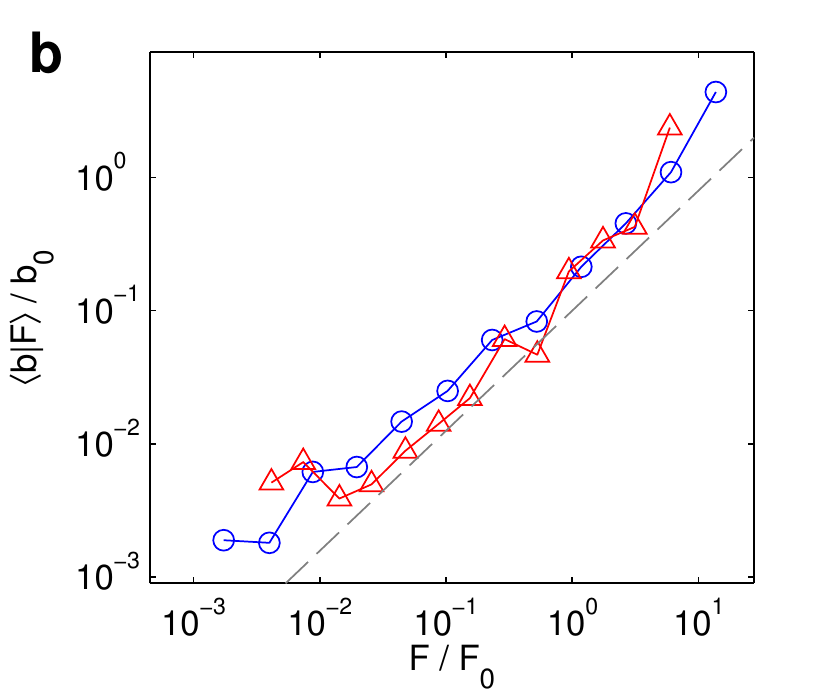}\hfill{}\includegraphics[width=0.32\columnwidth]{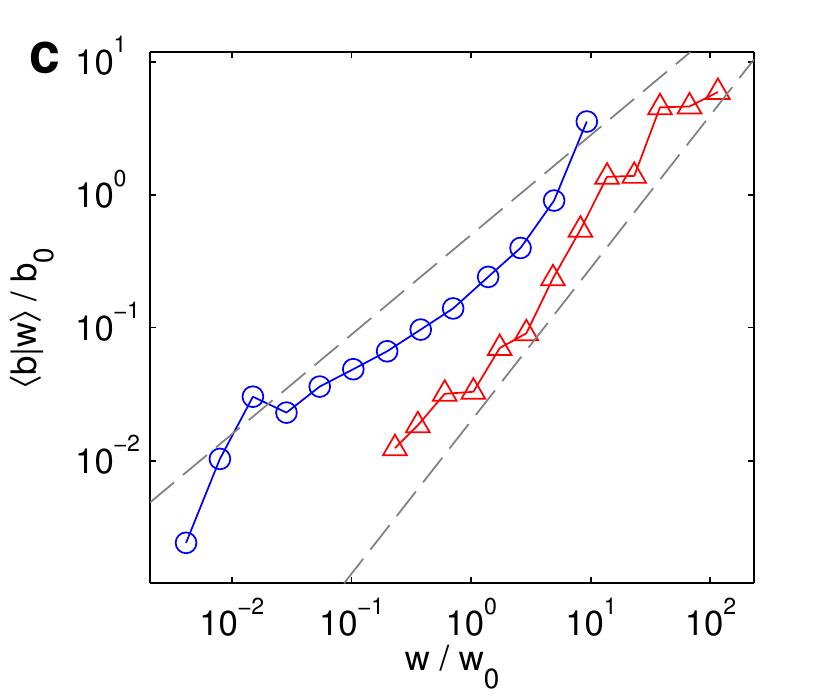}
\par\end{centering}

\caption{\label{fig:Correlations}Correlation structure of centrality measures.
\textbf{a:} The conditional mean degree $\left\langle k|F\right\rangle $
as a function of flux $F$. Both WAN (blue) and GCSN (red) exhibit
a similar sub-linear scaling with exponent $\eta\approx0.75$ across
the entire flux range. \textbf{b:} Conditional mean node betweenness
$\left\langle b|F\right\rangle $ as a function of $F$ is almost
identical in both networks with almost linear scaling $\zeta\approx0.9$.
\textbf{c:} Conditional weight betweenness $\left\langle b|w\right\rangle $
as a function of $w$ exhibits different scaling in both networks.
The WAN exhibits sub-linear scaling ($\xi\approx0.75$) contrary to
the GCSN for which super-linear scaling is observed ($\xi\approx1.15$).}
\end{figure}
Degree, flux and betweenness typically exhibit positive correlations
and scaling relationships with one another. For instance, recently-investigated
mobility networks~\cite{Barrat:2005p632,Brockmann:2008p1135,Kaluza:2010p5210}
exhibit a sub-linear scaling relation $k\sim F^{\eta}$ with exponent
$\eta\approx0.58$ and $\eta\approx0.7$. Figure~\ref{fig:Correlations}
compiles scaling relationships we observe in the WAN and GCSN. To
extract the scaling relationship we computed the mean of one centrality
measure $x$ conditioned on a second centrality measure $y$, that
is,\begin{equation}
\left\langle x|y\right\rangle =\frac{\int\text{d}x\, x\, p(x,y)}{\int\text{d}x\, p(x,y)}\label{eq:conditionalmean}\end{equation}
where $p(x,y)$ is the combined distribution of both. Our analysis
shows that both networks exhibit a sub-linear correlation of degree
with flux\begin{equation}
\left\langle k|F\right\rangle \sim F^{\eta}\label{eq:kofF}\end{equation}
with approximately identical exponent $\eta=0.75$ for both networks
and across 4 orders of magnitude of $F$. This is consistent with
previous findings and the intuitive notion that node connectivity
increases with traffic. A sub-linear scaling of degree with flux implies
that the typical weight $\left\langle w|F\right\rangle $ of links
connected to nodes of size $F$ scales according to \begin{equation}
\left\langle w|F\right\rangle \sim F/\left\langle k|F\right\rangle \sim F^{1-\eta}\label{eq:woff}\end{equation}
Since $\eta<1$ this implies that high flux nodes typically connect
to other nodes with stronger links, as expected for transportation
networks. The fact that $\eta$ is almost identical in both networks
is additional evidence that similar universal mechanism are responsible
for shaping the topological structure of both the WAN and GCSN. Similarly,
node betweenness scales as\begin{equation}
\left\langle b|F\right\rangle \sim F^{\zeta}\label{eq:bofF}\end{equation}
with an exponent $\zeta\approx1$ in both networks. A linear relationship
between node flux and betweenness can be explained by the heuristic
argument that typical betweenness values of a node increase linearly
with its degree $k$. Likewise, since shortest paths are computed
based on link weights, it is reasonable to assume that node betweenness
scales linearly with the typical link weight of a node and thus\begin{equation}
\left\langle b|F\right\rangle \sim\left\langle k|F\right\rangle \left\langle w|F\right\rangle \sim F^{\eta}F^{1-\eta}=F\label{eq:bofF2}\end{equation}
and hence one expects $\zeta\approx1$ as observed. Conditional mean
of link betweenness as a function of link weight $\left\langle b|w\right\rangle $
exhibit approximate scaling. Fig.~\ref{fig:Correlations}c suggests
sub-linear scaling for the WAN as opposed to super-linear scaling
for the GCSN. This difference in scaling in both networks is the first
marked difference that we observe in the statistics of centrality
measures. Possible explanations are the differences in overall connectivity
$\sigma$ in both networks (see Table~\ref{tab:networkproperties})
and that there is a statistically-significant difference in the way
that weights are distributed among the nodes.

\section{Network Resilience}

\label{sec:reilience}A key question in the context of large-scale
technological and infrastructural networks concerns their response
to local failure and resilience to accidental, partial breakdown or
anticipated attacks. Both the WAN and GCSN are subject to unpredictable,
recurrent, and extreme weather conditions that lead to repetitive
and regionally-localized failure that must be compensated for by re-routing
traffic or re-planning schedules.

\begin{figure}
\hfill{}\includegraphics[width=0.35\columnwidth]{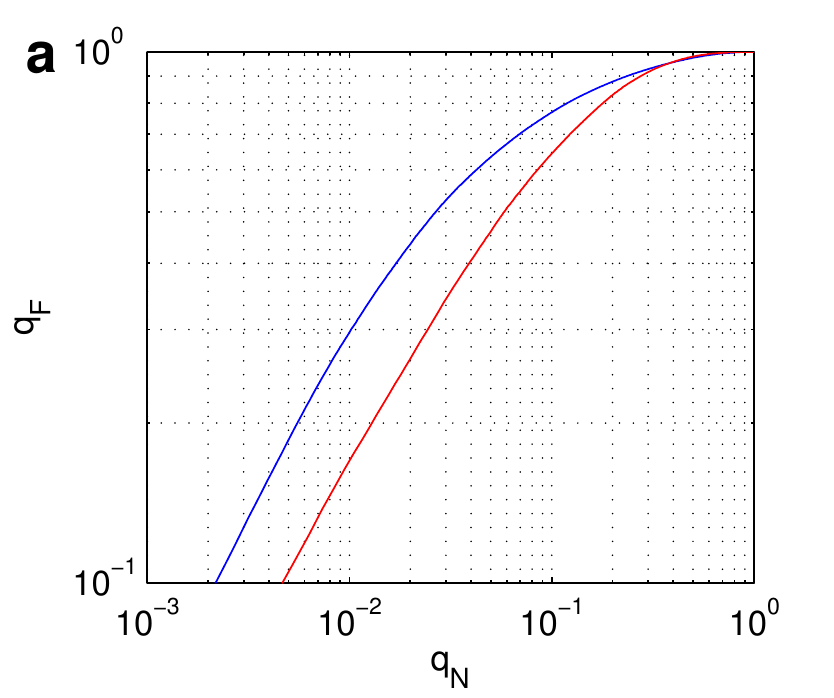}\hfill{}\includegraphics[width=0.35\columnwidth]{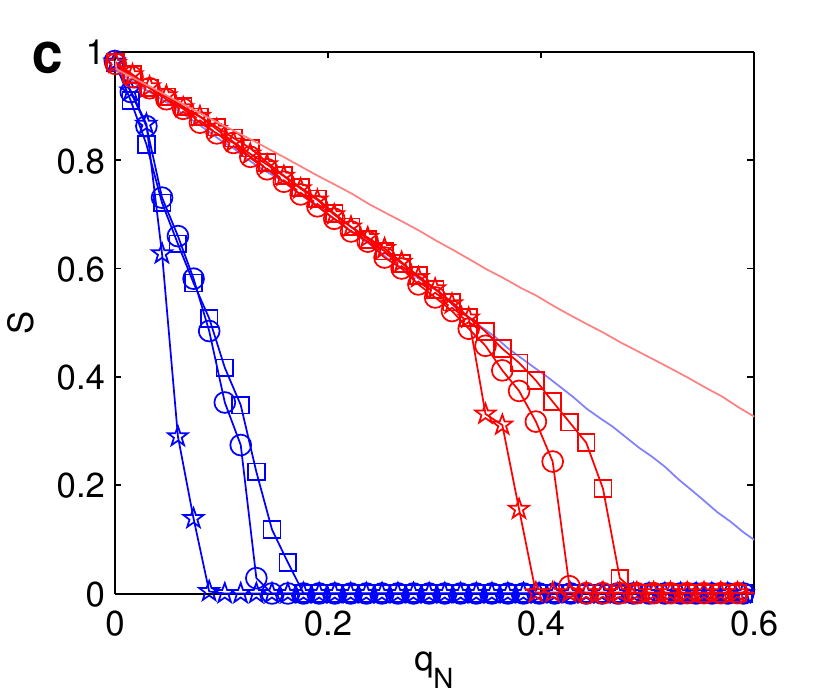}\hfill{}

\hfill{}\includegraphics[width=0.35\columnwidth]{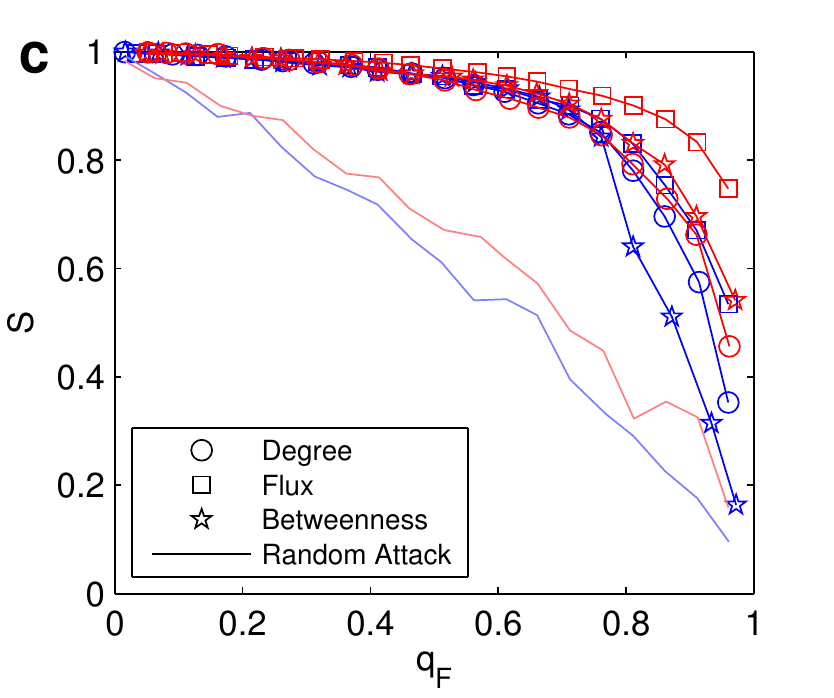}\hfill{}\includegraphics[width=0.35\columnwidth]{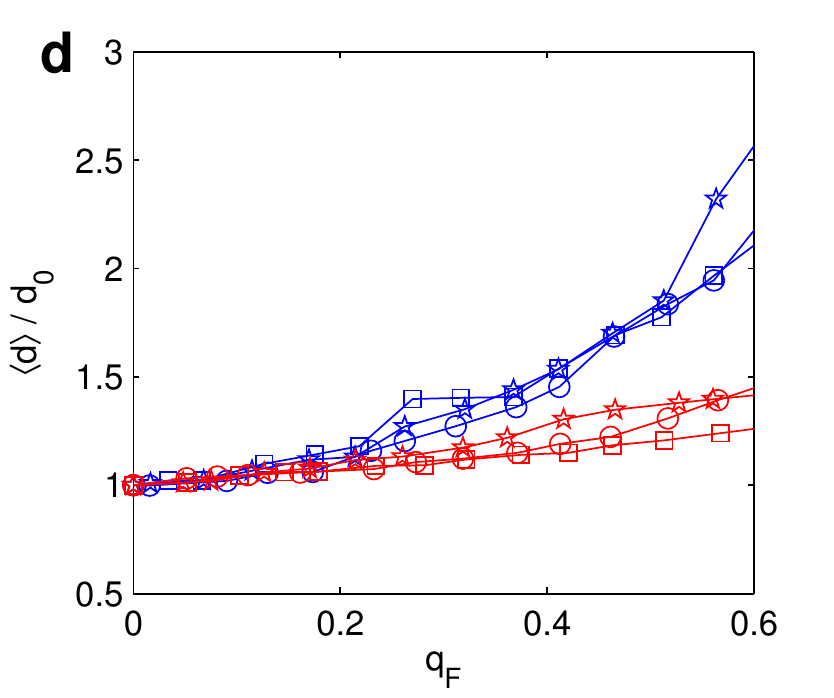}\hfill{}

\caption{\label{fig:resilience}Resilience properties of WAN (blue) and GCSN
(red) in response to random and selected node removal: \textbf{a)}
Fraction $q_{F}$ of total traffic that is carried by a fraction $q_{N}$
of top-flux nodes. In both networks a few most central nodes carry
substantial amount of traffic in the network.\textbf{ b)} The relative
size of the largest connected component of the networks as a function
of the fraction $q_{N}$ of removed nodes. Nodes are removed according
to the rank in terms of degree $k$ (circles), flux $F$ (squares),
betweenness $b$ (stars), or randomly (no symbols). Both networks
exhibit percolation thresholds under the attack protocols. Numerical
values of the percolation thresholds are given in the text. \textbf{c)}
The largest component as a function of fraction of removed traffic
does not exhibit a clear percolation threshold even when almost all
the traffic is removed. This indicates that the integrity of the entire
network is not altered substantially even if an unrealistically large
amount of traffic is reduced. \textbf{d)} The response to selected
node removal as reflected in network inflation. The panel depicts
the diameter of the network as defined by the mean shortest path between
all pairs of nodes. Both networks show a substantial increase in diameter
as a function of reduced traffic. The effect is more substantial in
the WAN.}
\end{figure}
Random failures and targeted attacks are typically investigated using
the framework of percolation theory~\cite{Cohen:2000p977,Cohen:2003p4342}.
A random (failure) or selected (attack) fraction $q$ of nodes is
removed from the network and structural responses of the network are
investigated as a function of $q$. Important insight was gained in
studies that investigated random or selected node removal in random
networks~\cite{Albert:2000p905,Chen:2008p197,Cohen:2000p977,Cohen:2003p4342}.
One of the most important findings of these studies was that scale-free
networks with power-law degree distributions respond strikingly differently
in scenarios that reflect random failures as opposed to selected removal
of central nodes. For instance, scale-free networks are relatively
immune to random removal of nodes and extremely sensitive to targeted
removal of high centrality nodes. Since centrality measures such as
degree, betweenness, and flux typically correlate in these networks,
this effectively amounts to removal of nodes that function as hubs.
One of the essential questions in this context addresses the critical
fraction $q$ of removed nodes that are required to disintegrate the
global connectivity of the network. This critical value is the percolation
threshold $q_{c}$: for $q<q_{c}$ the size of the giant component
(the largest subset of nodes that are connected by paths) is typically
the size of the entire network. Beyond the percolation threshold ($q>q_{c}$)
the networks falls apart into a family of disconnected, fragmented
sub-networks.

The resilience properties of the WAN and GCSN to sequential node removal
are depicted in Fig.~\ref{fig:resilience}. For each centrality measure
(degree, betweenness, and flux), we remove fractions $q$ of nodes
according to their rank with respect to $k$, $b$, and $F$, respectively.
We compare two different removal protocols. Since both networks are
strongly inhomogeneous, removing a fraction of nodes is not equivalent
to removing a fraction of traffic (see Fig.~\ref{fig:resilience}a).
For example $1\%$ of the most connected nodes account for $29.7\%$
of the entire traffic in the WAN and $17.6\%$ in the GCSN, and removal
of $10\%$ of nodes with highest flux is equivalent to reducing the
total traffic in the WAN by $76.9\%$ and in the GCSN by $64.5\%$.
Because of this pronounced nonlinear relationship, we compare resilience
of the network as a function of the fraction of removed nodes $q_{N}$
as well as the fraction of removed traffic $q_{F}$.

Figure~\ref{fig:resilience}b depicts the relative size of the giant
component $S$ as a function of $q_{N}$. Both networks are resilient
to random failures (we find that the giant component decreases lineary
with the fraction of nodes removed, i.e. $S\approx1-q_{N}$), although
the WAN is taking some excess damage from random failures. Furthermore,
we observe a percolation threshold for the targeted attacks in both
networks. The WAN exhibits a percolation threshold at $q_{N}^{c}=13.8\%,\,17.2\%$
and $9.4\%$, for node removal according to degree, flux and betweenness.
The thresholds are significantly larger for the GCSN at $q_{N}^{c}=44.3\%,49.0\%$
and $39.5\%$. In each network the threshold depends only weakly on
the choice of centrality measure because of the strong correlation
among different centrality measures. Note however that both networks
are most susceptible to removal according to betweenness rank, followed
by degree and node flux. The overall higher threshold in the GCSN
is caused by the greater connectivity $\sigma$ and mean degree $\left\langle k\right\rangle $
of the network (see Table~\ref{tab:networkproperties}).

Figure~\ref{fig:resilience}c depicts $S$ as a function of $q_{F}.$
The random failures appear to be more effective here because they
remove more nodes for a given fraction of removed traffic than the
targeted attacks, since not only high-centrality nodes are selected.
However, due to the strong nonlinear relationship between $q_{N}$
and $q_{F}$ it is evident that both networks are strongly resilient
to targeted attacks. Even substantial traffic reduction has virtually
no impact on the relative size of the giant component, for instance
when 50\% of the entire traffic is reduced in both networks, the giant
component is still larger than $90\%$ of the original network and
no percolation threshold is observed in the range up to $80\%$ of
traffic reduction. These traffic reductions are unrealistic when compared
to actual perturbations of real transportation networks. Percolation
thresholds are therefore never reached under realistic conditions.
Another approach that has been applied in unweighted networks~\cite{Albert:2000p905}
is based on the diameter of the network and its response to network
disruption. Typically when high-centrality nodes are removed from
the network, the diameter of the network increases as the shortest
paths connecting two arbitrary nodes lengthen due to the increasing
lack of hubs that can serve as connecting junctions. Figure~\ref{fig:resilience}d
shows that this inflation of the network in response to node removal
is observed in both networks. This effect is relatively independent
of the choice of centrality measure used in the removal protocol.
Furthermore, the GCSN is more robust to node reduction, which we believe
to be a consequence of the high connectivity of the network.

Both percolation analysis and network inflation have only limited
applicability in real world scenarios. Since real world network disruptions
never reach the percolation threshold and network inflation only address
global structural changes in network properties, a more refined quantity
is needed that can determine the response to external perturbations
below the percolation threshold and on a node by node basis. In section~\ref{sec:reilience}
we propose a technique to quantify network resilience that permits
the study of network pertubations in a more refined framework and
well below the percolation threshold based on shortest-path trees.
The key idea behind this technique is the ability to quantify the
effect of network disruptions for each node and perform network-wide
statistics of the impact of external pertubations or network disruptions.

\section{Network shortest-path trees}

\label{sec:SPD_SPTD}%
\begin{figure}
\includegraphics[width=1\columnwidth]{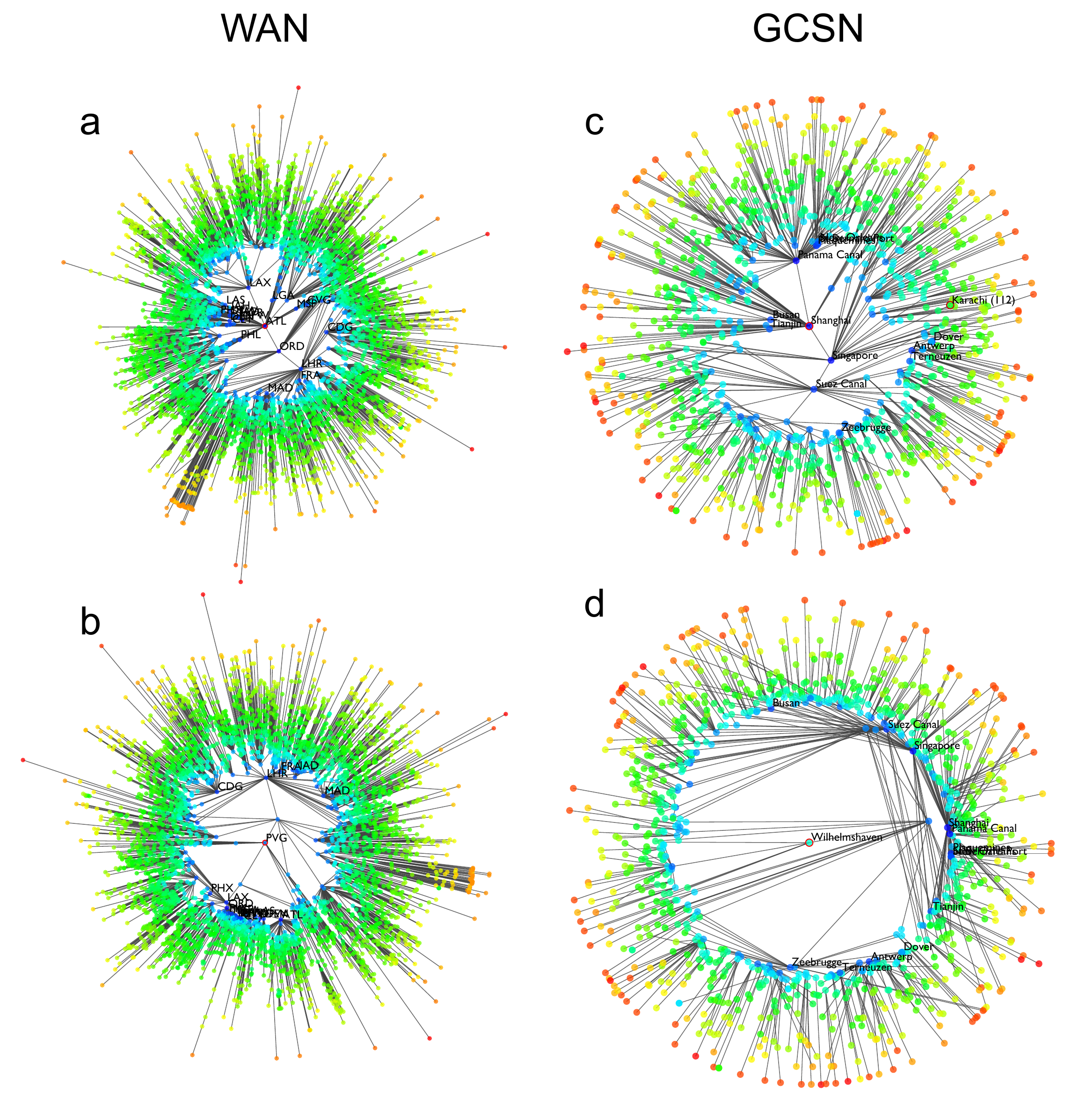}

\caption{\label{fig:Spato} Shortest-path tree structures and effective distances
in WAN and GCSN. \textbf{Left:} The panels depict the shortest-path
trees of airports ATL (Atlanta) and PVG (Shanghai). The radial distance
of the remaining airports with respect to these root nodes represent
the logarithm of the shortest-path distance to the reference node.
\textbf{Right:} The panels depict the GCSN shortest-path tree for
ports Wilhelmshaven, Germany and Shanghai. Note that the overall structure
of both representations is different, yet both networks share the
feature of circular arrangement according to node flux, encoded by
color (blue represents large flux nodes and orange small node flux).
Note that irrespective of the chosen root node, the closest nodes
in terms of effective distance are always high flux nodes and small
flux nodes are always peripheral in this representation. }
\end{figure}
Global properties of strongly heterogeneous, multi-scale networks,
such as connectivity, clustering coefficient, and diameter, as well
as statistical distributions of centrality measures, provide important
insight and may serve as quantitative classifiers for networks. However,
they cannot resolve properties and structures on a local scale. On
the other hand, local measures such as a node's individual degree,
betweenness centrality, or mean link weight of its connections provide
local information only and cannot capture global properties. Transportation
networks exhibit important structure on intermediate scales, so it
is vital to understand structural properties that are neither local
nor global in these networks. One way to approach this is to analyse
and investigate the structure of the entire network from the perspective
of a chosen node. Clearly, geographic distance is an important parameter
in this context as operation costs typically scale with geographic
distance. However, in complex multi-scale transportation networks
such as the WAN and the GCSN, geographic distance is rarely a good
indicator of the effective distance of connected nodes. High-flux
hubs in each network are typically connected by strong traffic bonds
even across very large geographic distances while smaller-flux nodes
can be connected by weak links although they may be geographically
close. A spatial representation as depicted in Fig.~\ref{fig:thenetworks}
is therefore a misleading way to convey effective distances in these
networks. 

An alternative representation can be obtained based on the notion
of proximity defined by Eq.~(\ref{eq:proximity}) and effective shortest
paths, Eq.~(\ref{eq:length of path}). Based on this notion we compute
the shortest paths of a chosen root node $i$ to all other nodes $j$.
The collection of links contributing to these paths form a shortest-path
tree $\mathbf{T}_{i}$ rooted at $i$. Spatial representations of
such trees are depicted in Fig.~\ref{fig:Spato} for each network
and two different root nodes. The radial distance in these figures
represents the effective, shortest-path distance $d_{ji}$. The lines
represent the connections of $\mathbf{T}_{i}$. Note that, although
the trees differ in both networks and for different root nodes, high-centrality
nodes tend to exhibit the smallest effective (shortest-path) distance
to the root node. Note also that the geometry of the networks exhibits
significant structural differences in both networks: In the WAN the
spatial distribution in the new representation is less regular and
the scatter in effective distance is larger than in the GCSN where
nodes reside in a well defined annular region.

\begin{figure}
\includegraphics[width=0.33\columnwidth]{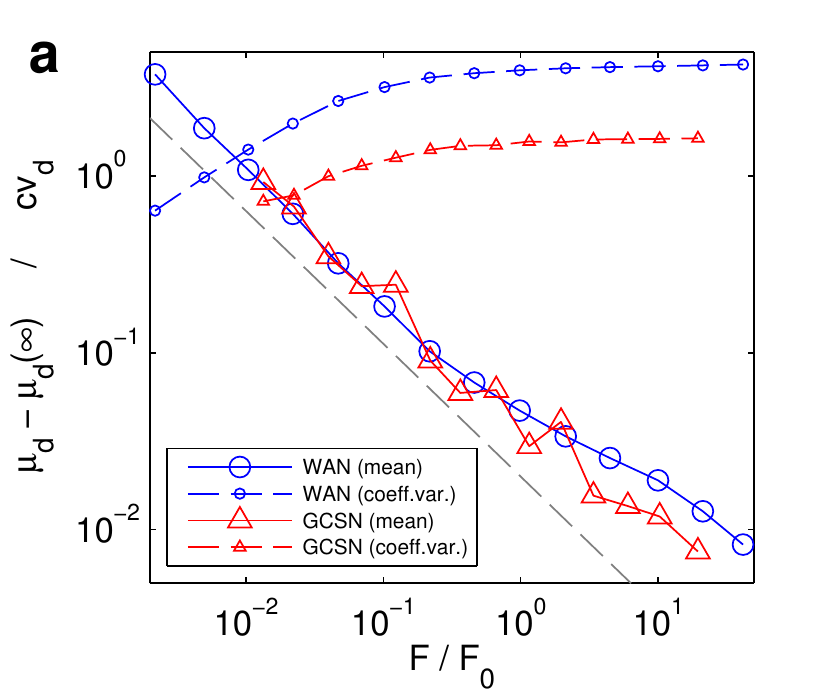}\hfill{}\includegraphics[width=0.33\columnwidth]{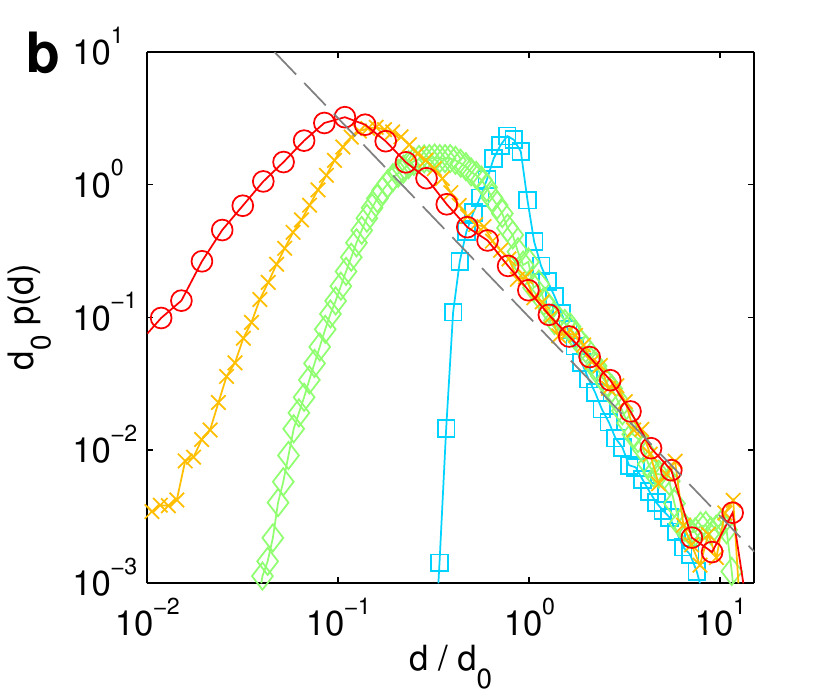}\hfill{}\includegraphics[width=0.33\columnwidth]{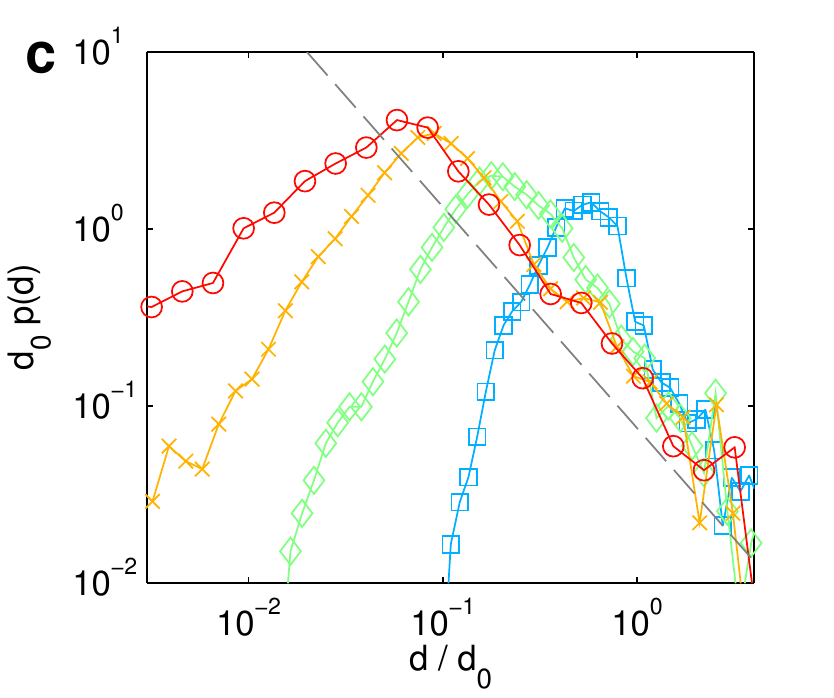}

\caption{\label{fig:pdfspd}Shortest-path distance statistics. \textbf{a:}
Conditional mean shortest path (Eq.~\ref{eq:muofF}) as a function
of node flux as well as the conditional coefficient of variation in
shortest paths. Both networks exhibit a universal decrease of expected
shortest path with node flux determined by an exponent $\tau\approx0.75$,
see Eq.~\ref{eq:muoff2}. The variability of shortest-path distance
increases in both networks, but less markedly in the GCSN than in
the WAN. \textbf{b:} The conditional distributions of shortest paths
for four subtypes of root nodes (WAN) ranked according to flux, where
red markers denote high flux and light blue denote low flux. \textbf{c:}
Same as in \textbf{b} for the GCSN. The dashed grey lines indicate
a scaling relation with exponent $\theta\approx1.5$ (WAN) or $\theta\approx1.25$
(GCSN).}
\end{figure}
In order to understand these qualitative differences and similarities
we investigate the distribution of the shortest-path distances conditioned
on the type of root node. The results of this analysis are depicted
in Fig.~\ref{fig:pdfspd}. Conditioned on the flux of the root node,
we compute the distribution of shortest-path distance, that is, $p(d|F).$
Based on this distribution we determine the expected distance of the
network from a node with specified flux as\begin{equation}
\mu_{d}(F)=\left\langle d|F\right\rangle \label{eq:muofF}\end{equation}
as well as the conditional coefficient of variation: \[
\text{cv}_{d}(F)=\frac{\sqrt{\left\langle d^{2}|F\right\rangle -\left\langle d|F\right\rangle ^{2}}}{\left\langle d|F\right\rangle }\]
The quantity $\mu_{d}(F)$ measures the typical distance from a root
node with flux $F$ to the rest of the network. The coefficient of
variation measures the statistical variability in $d$. Figure~\ref{fig:pdfspd}
depicts both quantities for the WAN and GCSN. Note that $\mu_{d}(F)$
behaves identically for both networks and can described by\begin{equation}
\mu_{d}(F)-\mu_{d}(\infty)\sim(F/F_{0})^{-\tau}\label{eq:muoff2}\end{equation}
with $\tau\approx0.75$. Note that this relation indicates the existance
of a lower limit to the typical effective distance for increasing
node flux $\mu_{d}(\infty)>0$ which implies that even extremely large
hubs exhibit a least distance to the rest of the network. Eq.~(\ref{eq:muoff2})
implies that mean effective distance decreases in a systematic way
with node centrality and according to the same relation in both networks.
However, the coefficient of variation increases monotonically with
$F$, which implies that the variability in effective distance increases
with the centrality of the root node. This can also be observed in
Fig~\ref{fig:pdfspd}b and c, which depicts the entire distribution
$p(d|F)$ for four categories of root nodes of different centrality.
For most central nodes $p(d|F)$ increases steeply for small values
of $d$ and exhibits an algebraic decay for large distance. As $F$
decreases, $p(d|F)$ attains a sharper peak as small distances disappear
from the distribution. This qualitative behavior is observed in both
networks. The asymptotic behavior for large effective distances is
approximately\[
p(d|F)\sim d^{-\theta}\]
with $\theta\approx1.5$ for the WAN and $\theta\approx1.25$ for
the GCSN.

\begin{figure}
\includegraphics[width=0.33\columnwidth]{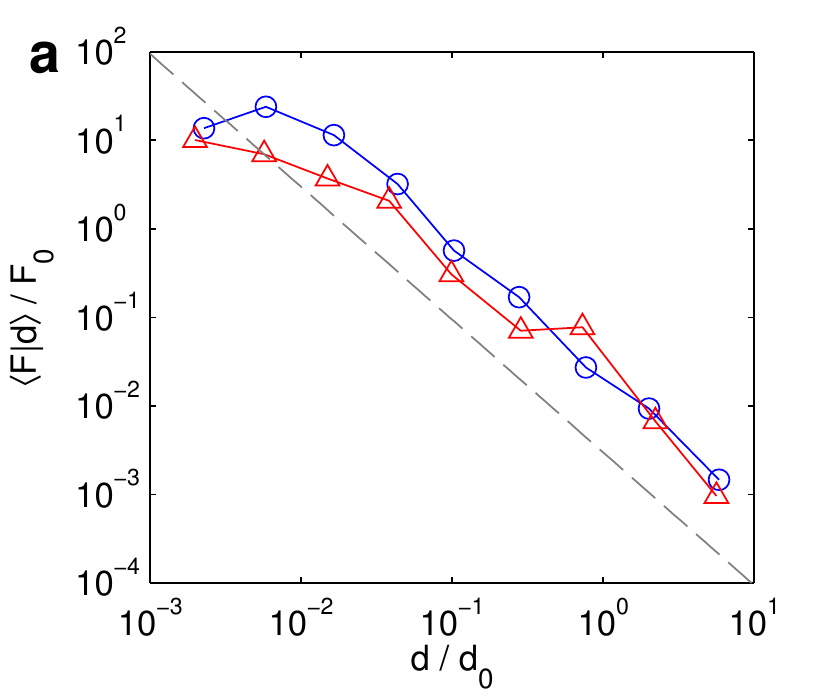}\hfill{}\includegraphics[width=0.33\columnwidth]{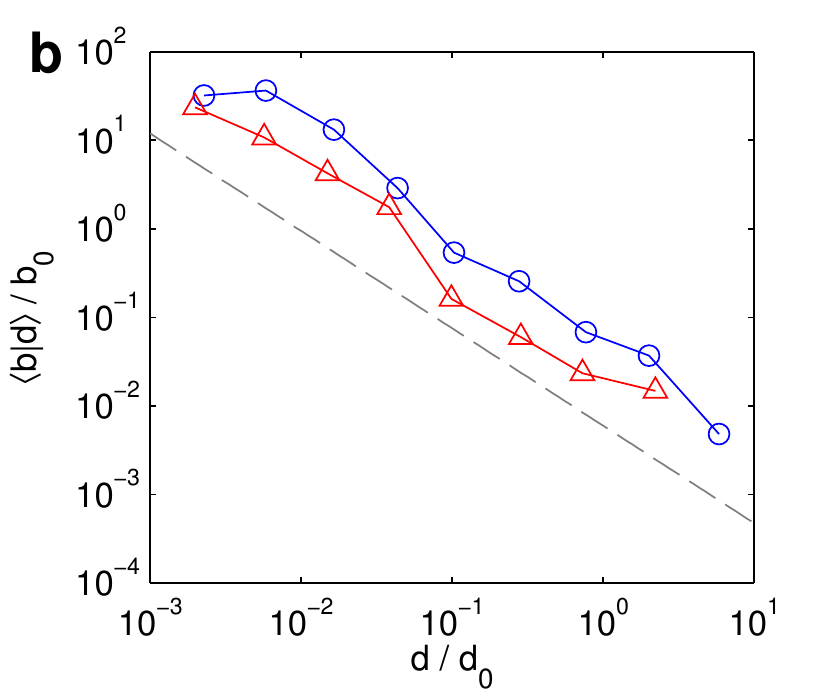}\hfill{}\includegraphics[width=0.33\columnwidth]{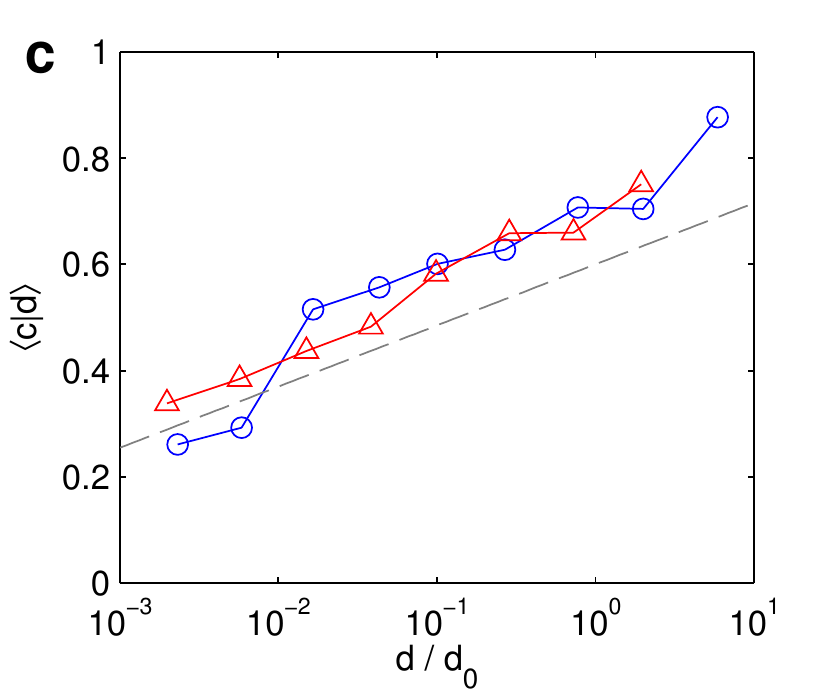}

\caption{\label{fig:pofxconditionedonr}Correlation analysis of shortest paths
and centrality measures (\textbf{a}) flux, (\textbf{b}) weighted node
betweenness for WAN (blue) and GCSN (red). Both networks exhibit almost
identical scaling relations. The scaling exponents are given by $\omega_{F}=1.5$,
$\omega_{b}=1.1$. \textbf{c:} the local clustering coefficient $c$
as a function of effective distance $d$ increases logarithmically. }
\end{figure}
A characteristic property of the network representation in Fig.~\ref{fig:Spato}
is that regardless of the properties of the root node, the rest of
the nodes tend to sort in concentric circles (effective distances)
according to centrality measures. A key question is then how effective
distance correlates with centrality measures. If there is a strong
correlation between effective distance and node centrality measures,
this implies that centrality measures dominate the placement of a
node in a network. 

In order to determine the relationship between effective distance
and centrality measures, we selected $2.5\%$ of the most central
nodes, according to degree, flux and betweenness and collected them
in a subset of nodes $\Omega$. The fraction of nodes in this set
represents $5\%$ of the entire network. The remaining $95\%$ of
the nodes are denoted by $\bar{\Omega}$. Based on this subset we
determine the distribution $p(x,d|\Omega)$, the probability of finding
a node in $\bar{\Omega}$ with centrality measure $x$ (degree, flux,
betweenness) and effective distance $d$ to the root nodes in $\Omega$.
From this we computed the conditional mean\begin{equation}
\left\langle x|d\right\rangle =\int x\, p(x|d,\Omega)=\int x\, p(x,d|\Omega)/p(d|\Omega),\label{eq:conditionalcentrality}\end{equation}
Figure~\ref{fig:pofxconditionedonr} depicts $ $$\left\langle F|d\right\rangle $
and $\left\langle b|d\right\rangle $ for both networks. Despite their
difference, WAN and GCSN exhibit almost identical scaling relations\begin{equation}
\left\langle F|d,\Omega\right\rangle \sim d^{-\omega_{F}}\quad\text{and}\quad\left\langle b|d,\Omega\right\rangle \sim d^{-\omega_{b}}\label{eq:scaling relations conditional centrality}\end{equation}
with $\omega_{F}\approx1.5$ and $\omega_{b}\approx1.1$, consistent
with the intuitive notion that centrality decreases with increasing
effective distance from central root nodes. Figure~\ref{fig:pofxconditionedonr}
also shows that the local clustering coefficient as a function of
$d$ approximately scales according to\begin{equation}
\left\langle c|d,\Omega\right\rangle \sim\log(d/d_{0}).\label{eq:clustering coefficient}\end{equation}
The logarithmic increase of the clustering coefficient implies that
in their peripheral regions the WAN and GCSN become less tree-like.
A plausible explanation is that low-centrality nodes that are connected
to the root nodes in $\Omega$ do not exhibit large fractions of connections
among one another, which indicates that high-centrality root nodes
function as {}``feed-in'' hubs to low centrality nodes. 

\begin{figure}
\includegraphics[width=0.33\columnwidth]{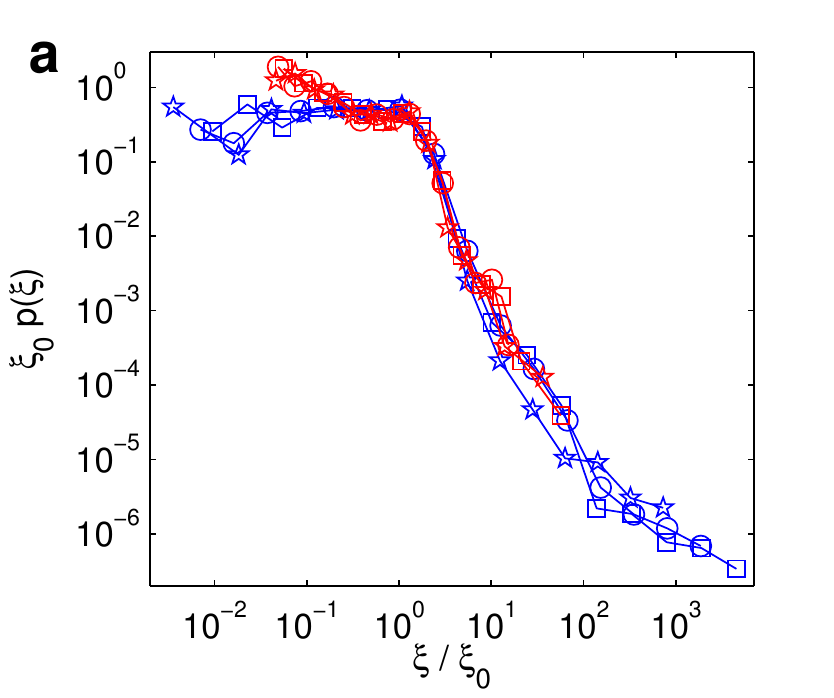}\hfill{}\includegraphics[width=0.33\columnwidth]{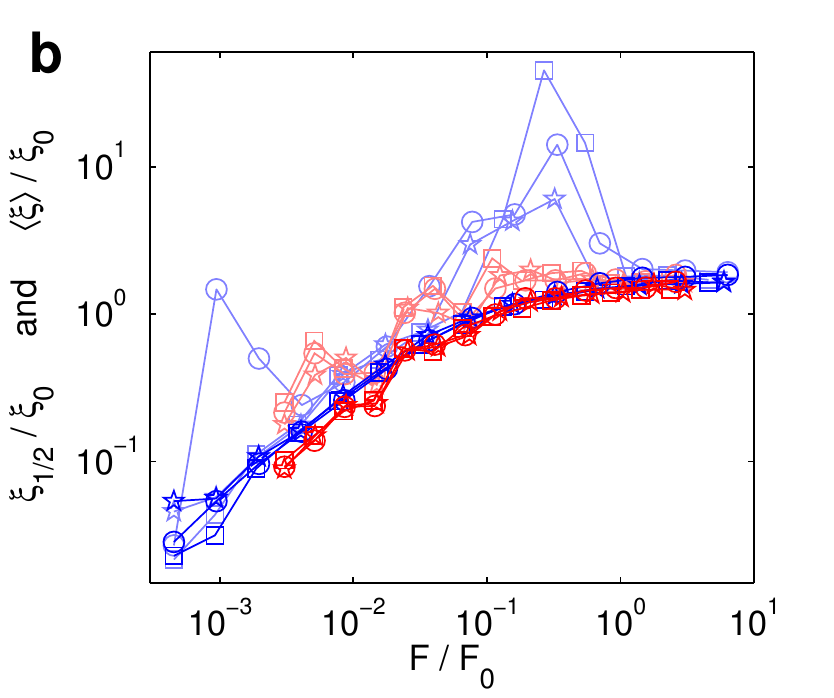}\hfill{}\includegraphics[width=0.33\columnwidth]{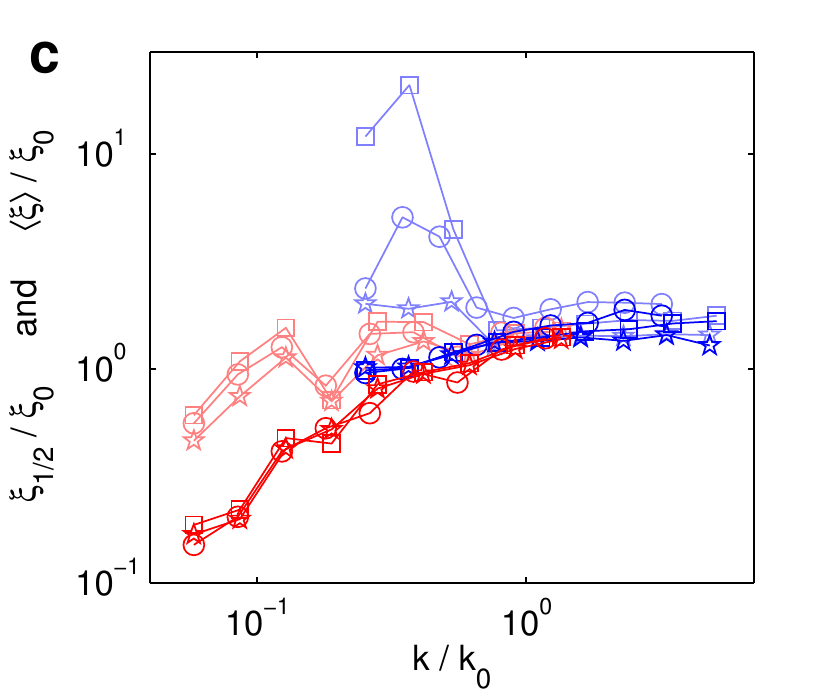}

\caption{\label{fig:impactfactor} The distribution of impact in response to
the removal of central nodes of the system as determined by degree
(circles), flux (squares), and betweenness (stars) in the WAN (blue)
and GCSN (red). \textbf{a:} $p(\xi)$ is the probability distribution
of impact factors computed for each node in response to the various
attacks. Note that apart from a scaling factor the distribution of
impact is identical in all three attack scenarios and both networks.
The impact factor $\xi$ ranges over many orders of magnitude. \textbf{b:}
Universal behavior is also observed in the dependence of impact factor
as a function of node flux $F$. The y-axis measures the normalized
median impact ($\xi_{1/2}/\xi_{0}$, solid lines) and normalized mean
impact ($\left\langle \xi\right\rangle /\xi_{0}$, faint lines). \textbf{c:}
The dependence of impact factor on node degree.}
\end{figure}

\subsection{Resilience and shortest paths}

The concept of shortest-path trees can also give insight into the
networks' resilience properties discussed in section~\ref{sec:reilience}.
In response to removal of a fraction $q_{N}$ of most central nodes
or the equivalent fraction of traffic $q_{F}$ in the entire network,
we can compute the impact by investigating the change of shortest-path
trees $\mathbf{T}_{i}$ for each root node $i$, that is, we can quantify
the impact of the network disruption from the perspective of every
node. To this end we define a node's impact factor as \begin{equation}
\xi_{i}=\frac{\Delta\bar{d}_{i}}{\bar{d}_{i}}\label{eq:imapct factor}\end{equation}
where $\bar{d}_{i}$ is the median shortest-path distance from reference
node $i$ to all other nodes $j$, and $\Delta\bar{d}_{i}$ the change
of this median in response to the network disruption. This impact
factor is different for every node and the distribution $p(\xi)$
gives insight into the variability of how individual nodes are affected
by the network disruption~\cite{volcano}. Figure~\ref{fig:impactfactor}a
illustrates $p(\xi)$ for scenarios in which the entire traffic was
reduced by $30\%$ through the removal of high-centrality nodes. The
distribution $p(\xi)$ is independent of the measure of centrality
and also identical in both networks. Below a typical impact of $\xi_{0}$
the distribution of impact factors $p(\xi)$ is uniform and for $\xi>\xi_{0}$
it decreases slowly, ranging over many orders of magnitude. A question
that immediately arises is what nodes in the network experience the
largest impact. Figures~\ref{fig:impactfactor}b/c depict the mean
$\left\langle \xi\right\rangle $ and median $\xi_{1/2}$ conditioned
on the flux $F$ and degree $k$. Both the WAN and GCSN exhibit the
same dependence, with increasing centrality, the median impact increases
monotonically and reaches the typical asymptotic value $\xi_{0}.$
However, the mean $\left\langle \xi\right\rangle $ as a function
of $F$ exhibits strong fluctuations for intermediate ranges of $F$.
The explanation for this phenomenon is that the relative impact for
low centrality nodes is small because $\bar{d}$ in the unperturbed
network is very large. Nodes of intermediate centrality are affected
strongly because their mean effective distance to the network $\bar{d}$
is of intermediate size as they primarily connect to hubs in the network
by strong links. When the hubs are removed from the network, these
nodes experience a strong increase in impact as $\Delta\bar{d}$ is
increased substantially. A similar effect is seen in the behavior
of $\left\langle \xi\right\rangle $ as a function of degree $k$.

\section{Discussion}

\label{sec:discussion}

The comparative analysis of the worldwide air-transportation network
and the global cargoship network presented here is a first step towards
a better understanding of the organizational structure, the evolution
and management of large scale infrastructural networks in general.
The statistical analysis of node and link centrality measures and
their correlations revealed a suprising degree of similarity of both
networks despite their different purpose, scale and connectivity.
We believe that this is strong evidence for common underlying principles
that govern the growth and evolution of infrastructural networks.
This is also supported by the variety of simple algebraic scaling
relations that we extracted from both networks.

Our analysis revealed an unusual discontinuity in the distribution
of both link and node betweenness. This suggests that strongly heterogeneous
transportation and mobility networks exhibit a natural functional
separation of links and nodes into two distinct groups. Interestingly,
this discontinuity is localized at the same relative betweenness value
and has approximately the same magnitude in both networks. We conclude
that this natural separation into different classes of nodes and links
might well be a universal feature of these transportation networks
as well and could be a starting point for further investigations along
these lines.

The analysis of network resilience showed that because of their dense
connectivity, both networks cannot be investigated by conventional
analysis techniques such as percolation theory or network diameter
inflation. The percolation threshold for both networks lies well beyond
any realistic network pertubations. The alternative approach based
on effective distance, shortest paths, and shortest-path trees allows
a better, more intuitive representation of networks and resilience
analysis, taking into account the fact that nodes that are connected
by strong traffic are effectively closer than nodes that are connected
by weak links and investigating network pertubations from the viewpoint
of chosen reference nodes. Furthermore, the shortest-path-tree representation
revealed an interesting correlation of effective shortest path distance
and node centrality measures such as flux, degree, and betweenness
and an interesting symmetry in both networks: On average, any node
in the network is closest to the subset of nodes with high centrality.
This has fundamental implications for spreading phenomena on these
types of networks. Whereas global disease dynamics, for example, is
characterized by highly complex spatio-temporal patterns when visualized
in conventional geographical coordinates, we expect these patterns
become simpler and thus better understood when shortest-path-tree
representations are employed. Since the shortest-path-tree representations
are structurally similar in both networks one might expect a strong
dynamic similarity of otherwise unrelated spreading phenomena that
occur in these networks, for example the global spread of emergent
human infectious diseases on the worldwide air-transportation network
and human mediated bioinvasion processes on the global cargoship network.
We conclude that our results can serve as a starting point for both
the development of theories for the evolution of large scale transportation
networks and dynamical processes that evolve on them.
\begin{acknowledgments}
The authors wish to acknowledge support from the Volkswagen Foundation.
\end{acknowledgments}
\bibliographystyle{apsrev}
\bibliography{bibliography}

\end{document}